\documentclass[a4paper]{spie}
\usepackage{amsmath,amssymb,graphicx,color}
\usepackage[version=3]{mhchem}
\usepackage{dcolumn}
\usepackage{bm}
\usepackage[mathlines]{lineno}
\usepackage{epsfig}
\usepackage{ulem}
\usepackage{graphicx}
\usepackage[cp1251]{inputenc}
\usepackage[english]{babel}
\usepackage{epstopdf}

\title{Towards all-dielectric metamaterials and nanophotonics}

\author{Alexander Krasnok\supit{a}, Sergey Makarov\supit{a}, Mikhail Petrov\supit{a}, Roman Savelev\supit{a}, Pavel Belov\supit{a},
and Yuri Kivshar\supit{b}
\skiplinehalf
\supit{a}ITMO University, St.~Petersburg 197101, Russia;\\
\supit{b}Nonlinear Physics Center and CUDOS@ANU, Research School Physics and Engineering,
Australian National University, Canberra ACT 0200, Australia}

\begin{document}
\maketitle

\begin{abstract}
We review a new, rapidly developing field of all-dielectric nanophotonics which allows to control both magnetic and electric
response of structured matter by engineering the Mie resonances in high-index dielectric nanoparticles. We discuss optical
properties of such dielectric nanoparticles, methods of their fabrication, and also recent advances in all-dielectric
metadevices including couple-resonator dielectric waveguides, nanoantennas, and metasurfaces.
\end{abstract}


\section{Introduction}

Modern technologies largely depend on the rapidly growing demands for powerful computational capacities and efficient information processing, so the development of conceptually new approaches and methods are extremely valuable.  One of those approached is based
on the use of light instead of electrons~\cite{CaulfieldNP2010}, replacing  electrons by photons as the main information carriers. The advantages of light for fast computing are obvious: parallel transfer and processing of signals using polarization and orbital momentum of photons as additional degrees of freedom~\cite{Willner_2014}, as well as a possibility of multi-frequency operations. However, one of the most important advantages of optical technology over electronics is its high operating frequency around 500 THz. However, photons as alternative information carriers have relatively large ``sizes'' determined by their wavelength. That is why they interact weakly with nanoscale (subwavelength) objects such as quantum emitters, subwavelength waveguides, and others. This problem call for challenges in creating and developing novel light control tools at the nanoscale.

\begin{figure}[!b] \centering
\includegraphics[width=0.6\columnwidth]{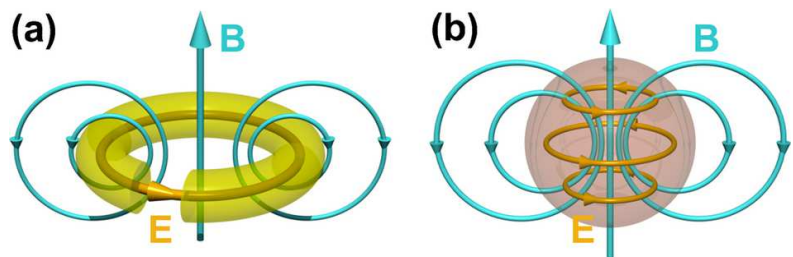}
\caption{Schematic representation of electric and magnetic field distribution inside (a) metallic split-ring resonator and (b)
high-refractive index dielectric nanoparticle at the frequency of the magnetic resonance.~\cite{KuznetsovSciRep2012}}
\label{fig:SRR+SiNP}
\end{figure}

An efficient light manipulation means simultaneous control of its electric and magnetic components. However, the magnetic response of
natural materials is very weak because of small electron's spin contribution at high frequency. This is the reason why photonic
devices operate mainly with the electric part of light wave. At the same time, magnetic dipoles are very common sources of magnetic
field in nature. The field of the magnetic dipole is usually calculated as the limit of a current loop shrinking to a point. The
fields configuration is equivalent to the one of an electric dipole considering that the electric and magnetic fields are exchanged.
The most common example of a magnetic dipole radiation is an electromagnetic wave produced by an excited metal split-ring resonator
(SRR), which is a basic constituting element of metamaterials (see Fig.~\ref{fig:SRR+SiNP}a)~\cite{Soukoulis, Pendry, Shelby}. The
real currents excited by external electromagnetic radiation and running inside the SRR produce a transverse oscillating up and down
magnetic field in the center of the ring, which simulates an oscillating magnetic dipole. The major interest of these artificial
systems is due to their ability to response to a magnetic component of incoming radiation and thus to have a non-unity or even
negative magnetic permeability ($\mu$) at optical frequencies, which does not exist in nature. This provides possibilities to design
unusual material properties such as negative refraction~\cite{Shalaev, Zheludev, Soukoulis, KivsharNM2012},
cloaking~\cite{Leonhardt}, or superlensing~\cite{Pendry00}. The SRR concept works very well for gigahertz~\cite{Smith, Shelby},
terahertz~\cite{Padilla06} and even near-infrared (few hundreds THz)~\cite{Liu08} frequencies. However, for shorter wavelengths and
in particular for visible spectral range this concept fails due to increasing losses and technological difficulties in fabrication of
smaller and smaller constituting split-ring elements~\cite{Soukoulis07}. Several other designs based on metal nanostructures have
been proposed to shift the magnetic resonance wavelength to the visible spectral range~\cite{Shalaev, Zheludev}. However, all of them
are suffering from losses inherent to metals at visible frequencies.

\begin{figure}[!t] \centering
\includegraphics[width=0.6\columnwidth]{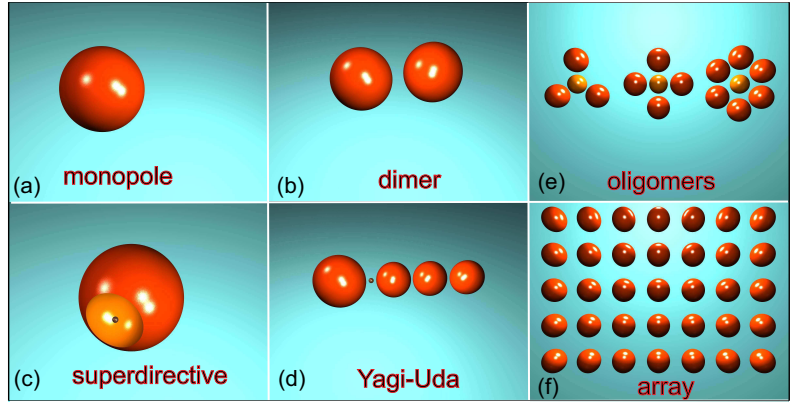}
\caption{The variety of photonics structures based on high-indexed dielectric nanoparticles: (a) single dielectric particle as a
monopole nanoantenna and nanoantenna ``Huygens source''~\cite{Schuller:OE:2009, Bonod10, Geffrin:NC:2012, Rigneault08,
KivsharAPL2014}; (b) dimer nanoantenna~\cite{BonodAPL2014, Samusev2015}; (c) superdirective dielectric
nanoantenna~\cite{KrasnokNanoscale}; (d) Yagi-Uda naoantenna~\cite{KrasnokOE2012, BonodOE2012}; (e) oligomer
nanostructures~\cite{Miroshnichenko:NL:2012}; (f) one-, two- and three-dimensional periodic arrays including discrete
waveguides~\cite{JunjiePRA2009, SavelevPRB2014}, metasurfaces and metamaterials~\cite{ShullerPRL2007, ShiAdvMat2012}.}
\label{fig:alldielnano}
\end{figure}

An alternative approach to achieve strong magnetic response with low losses is to use nanoparticles made of high-refractive index
dielectric materials~\cite{Grzegorczyk2007, ShullerPRL2007, Meng2008, Guizal2009, Zhao09, Cummer_08, Brener_12, Lukyanchuk13,
StaudeACSNano2013, Zhou2014, Habteyes2014}. As it follows from the exact Mie solution of light scattering by a spherical particle,
there is a particular parameter range where strong magnetic dipole resonance can be achieved. Remarkably, for the refractive indices
above a certain value there is a well-established hierarchy of magnetic and electric resonances. In contrast to plasmonic particles,
the first resonance of dielectric nanoparticles is a magnetic dipole one, and takes place when the wavelength of light inside the
particle equals to the diameter $\lambda/n_{s}\simeq2 R_{s}$, where $\lambda$ is a wavelength in a free space, $R_{s}$ and $n_{s}$
are the radius and refractive index of spherical particle. Under this condition the polarization of the electric field is
anti-parallel at opposite boundaries of the sphere, which gives rise to strong coupling to circulation displacement currents, while
magnetic field oscillates up and down in the middle (Fig.~\ref{fig:SRR+SiNP}b).

Several years ago it was theoretically predicted and experimentally proved~\cite{EvlyukhinPRB2010, KuznetsovSciRep2012,
EvlyukhinNL2012} that high-indexed dielectric particles can posses induced magnetic dipole moment, and, as opposed to plasmonic ones,
they do not have dissipative losses because of absence of free carriers. It is worth noting that the dielectric particle resonances
are not fixed. The resonant frequency can be controlled by changing the size and shape of the particle as well as the ambient
conditions~\cite{EvlyukhinPRB2011, EvlyukhinJOSAB2013, EvlyukhinSciRep2014}. Moreover, the electric and magnetic dipole resonances
can be overlapped in spectral range, bringing a number of unique optical properties. These advances of all-dielectric structures can
be a good alternative to plasmonic ones. This ''magnetic light'' concept opened the door to all-dielectric oligomer sensors and
nanoantennas~\cite{KrasnokOE2012, Miroshnichenko:NL:2012, Staude_2014Small, Filonov_oligomer, KrasnokNanoscale}, dielectric
waveguides~\cite{Savelev2014_1}, nonlinear optics~\cite{ShcherbakovNL2014}, all-dielectric Huygens’
metasurfaces~\cite{Staude_15}, and metamaterials~\cite{Cummer_08, Zhao09, Brener_12, ShiAdvMat2012, Valentine2014}.

This paper is devoted to review of photonic devices based on high-index dielectric nanoparticles and is organized as follows. In
Section~\ref{sec:OpticalProperties} the optical properties of dielectric particles with high refractive index are discussed. In
Section~\ref{sec:waveguides} we give the examples of the fabrication methods and produced structures. In
Section~\ref{sec:metamaterials} we review the studies on all-dielectric metamaterials and metasurfaces. In
Section~\ref{sec:waveguides} we review the studies on dielectric discrete waveguides. Section~\ref{sec:nanoantennas} contains
discussion on all-dielectric nanoantennas and oligomers, including ``Huygens source'' antennas [Fig.~\ref{fig:alldielnano}a],
Yagi-Uda [Fig.~\ref{fig:alldielnano}d], and superdirective nanoantennas with notch on its surface [Fig.~\ref{fig:alldielnano}c].

\section{Optical properties of high-index dielectric nanoparticles}
\label{sec:OpticalProperties}
\begin{figure}[!t] \centering
\includegraphics[width=0.5\columnwidth]{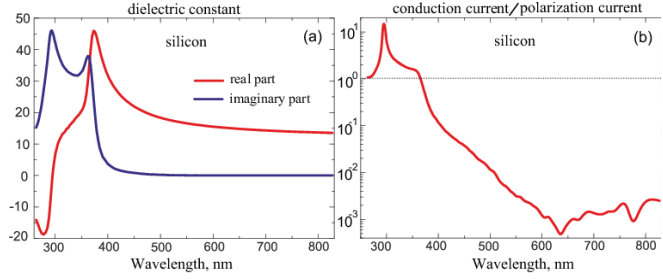}
\caption{(a) Real and imaginary parts of permittivity of crystalline silicon (Si)~\cite{VuyeSi}. (b) Ratio of conductivity current
and displacement current in crystalline silicon; Displacement current strongly exceeds conductivity current at wavelengths
appoximately above 400 nm -- in this spectral range pure silicon can be considered as dielectric.}
\label{fig:epsSi}
\end{figure}

It is known that two types of current appear as a sources of electromagnetic field in Maxwell equations: conductivity current and
displacement current~\cite{Novotny_Hecht_book}. In dielectrics and semiconductors, in spectral range far from their main absorption
band, displacement current strongly exceeds conductivity current. As an example, in Fig.~\ref{fig:epsSi}a spectral dependencies of
real and imaginary parts of refractive index, measured at a room temperature, are shown~\cite{VuyeSi}. In Fig.~\ref{fig:epsSi}b
ratio of conductivity current and displacement current is shown. One can see that displacement current strongly exceeds conductivity
current at wavelengths approximately above 400 nm. Therefore, from electrodynamics point of view pure silicon can be considered as a
dielectric in this spectral range.

Displacement current increases with the increase of permittivity of dielectric. It turns out that displacement current can be strongly
increased in dielectric nanoparticles large enough for emerging of Mie-resonances~\cite{Bohren}. Moreover, if such resonator is made
of dielectric with high permittivity (silicon, for example), dipole resonance conditions can be fulfilled for the particle with
subwavelength size.

Dielectric nanoparticle can be considered as an open resonator, that supports different types of electromagnetic field configuration
-- eigenmodes. Exact analysis of the diffraction of a plane wave by a spherical particle, known as Mie scattering, shows that
nanoparticle can support electric and magnetic eigenmodes of different order~\cite{Bohren}. The number of excited modes and their
order depends on the ratio $\gamma=\lambda/R$, where $\lambda$ is the wavelength of the incident radiation, and $R$ is the radius of
dielectric particle. If $\gamma$ is much more than unity, the particle is optically small and diffraction by such particle can be
described by Rayleigh approximation. When $\gamma$ decreases (for example when the wavelength decreases and radius of the particle is
constant), first \textit{magnetic type} dipole resonance is formed. Field lines at this resonance are shown in
Fig.~\ref{fig:SRR+SiNP}b. Scattering field of the particle corresponds to the scattering field of magnetic dipole in this case. With
further decreasing of $\gamma$, first electric type dipole resonance is formed. For even less values of $\gamma$ high order multipole
modes are excited.
\begin{figure}[!b] \centering
\includegraphics[width=0.5\columnwidth]{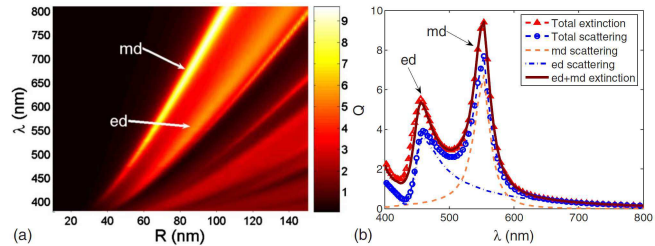}
\caption{(a) Scattering efficiency spectra of Si spherical particles with the radius R located in air. (b) Extinction and scattering
spectra of a Si particle ($R=65$nm). The arrows indicate the electric dipole (ed) and magnetic dipole (md) contributions to the total
efficiencies. From
the Ref.~\cite{EvlyukhinPRB2010}.}
\label{fig:SiNP_extinction}
\end{figure}

Resonance frequencies of a spherical particle can be determined from the condition:
\begin{equation} \label{eq_Mie_polarizabilities}
\mathrm{Re}(\alpha_e)=\mathrm{Re}(i\dfrac{3\varepsilon_h}{2k_h^3}a_1)=0 \text{,} \quad
\mathrm{Re}(\alpha_m)=\mathrm{Re}(i\dfrac{3}{2k_h^3}b_1)= 0,
\end{equation}
where $\alpha_e$ and $\alpha_m$ are electric and magnetic polarizabilities, respectively,$\varepsilon_h $ is the host permittivity,
$k_h=\sqrt{\varepsilon_h}\omega/c$ is the wavenumber of light in host media, $\omega$ is the angular frequency, $c$ is the speed of
light in vacuum, $a_1$ and $b_1$ are the scattering coefficients~\cite{Bohren}. In particular, in the Ref.~\cite{EvlyukhinPRB2010} it
was shown, that for silicon with permittivity about 18 in the visible range, conditions of the lowest order multipole (dipole) resonances are fulfilled for a spherical particle with radius $\approx70$ nm. Numerically calculated extinction and scattering spectra are shown in
Fig.~\ref{fig:SiNP_extinction}. Here, electric and magnetic dipole contributions are marked with $ed$ and $md$, respectively.
Resonance frequencies of the particle can be shifted not only by changing its size, by also its
shape~\cite{EvlyukhinPRB2011,EvlyukhinJOSAB2013,EvlyukhinSciRep2014}, and almost complete absence of conductivity currents in the silicon in optical frequency range leads to low dissipative losses, in contrast to plasmonics, where strong field localization always accompanied by high dissipation. So, exploiting such dielectric particles with magnetic response one can design different low-loss nanostructures, composite materials and metasurfaces with unique functionalities.

\section{Methods for the fabrication of nanoparticles}
\label{sec:prod}
\begin{figure}[!b] \centering
\includegraphics[width=0.5\columnwidth]{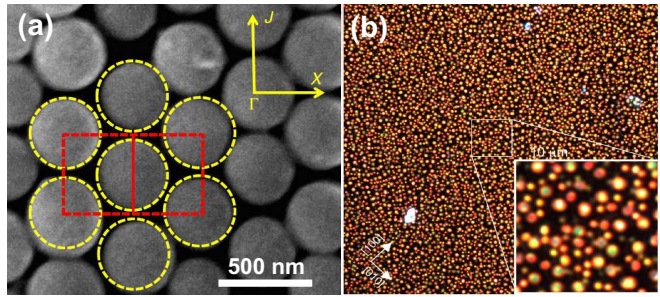}
\caption{Electron microscopy image of self-aligned silicon nanoparticles obtained by chemical deposition (a)~\cite{ShiNC2013}.
Dark-field optical image of silicon nanoparticles obtained by thin film dewetting (b)~\cite{AbbarchiACSNano2014}.}
\label{fig:Fabr_1}
\end{figure}

Silicon is the most frequently used high-index dielectric in optical and IR ranges owing to its relatively low cost and low imaginary
part of the refractive index. However, technology of fabrication of silicon nanoparticles with Mie-resonances has been developing
intensively during the last several years, resulting in emerging of various techniques. The proposed methods of silicon nano- and
microparticles fabrication can be classified on the level of the particles size and location controllability.

The fabrication method of silicon nanoparticles with different sizes can be carried out by means of chemical vapor deposition
technique, in which disilane gas (Si$_2$H$_6$) decomposes at high temperatures into solid silicon and hydrogen gas by the following
chemical reaction: Si$_2$H$_6 \rightarrow$ 2Si(s) + 3H$_2$(g). Spherical poly-crystalline silicon nanoparticles were produced by this
method~\cite{ShiAdvMat2012}. Further, fabrication of monodispersed silicon colloid was achieved via decomposition of trisilane
(Si$_3$H$_8$) in supercritical n-hexane at high temperature~\cite{ShiNC2013}. In this advanced method, the particles size can be
controlled by changing of trisilane concentration and temperature of the reaction. This relatively simple method allows one to obtain
plenty of similar silicon nanoparticles with size dispersion of several percents, which can be ordered into hexagonal lattice via a
self-assembly process [Fig.~\ref{fig:Fabr_1}b]. The main disadvantage of this method is the porosity and high hydrogen content in
each nanoparticle as well as necessity of their additional ordering to fabricate functional structure.

Disordered silicon nanoparticles of different sizes can be also produced via dewetting of thin supported silicon film after its
heating [Fig.~\ref{fig:Fabr_1}b]~\cite{AbbarchiACSNano2014}. In this case, the nanoparticles can be crystalline and their sides are
aligned along crystallographic facets. The main controlling parameters in this method are heating temperature and the film conditions
(defects and initial pattern)~\cite{AbbarchiACSNano2014}. In thin film dewetting technique the nanoparticles size and location
control can be achieved only by using additional lithographical methods, which is even more complicated in comparison with the
chemical deposition techniques. Indeed, both above mentioned methods are more suitable for high-throughput and low-cost nanoparticles
fabrication.
\begin{figure}[!t] \centering
\includegraphics[width=0.5\columnwidth]{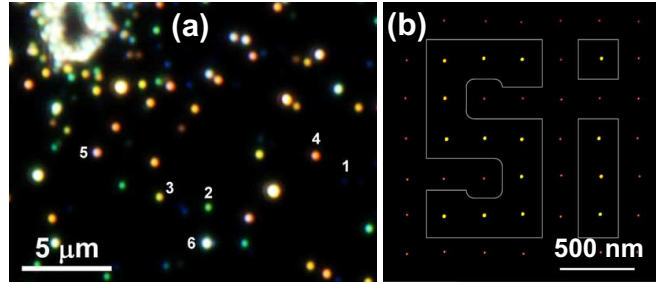}
\caption{Dark-field optical image of silicon nanoparticles obtained via femtosecond laser ablation of bulk silicon
(a)~\cite{KuznetsovSciRep2012} and thin silicon film(b)~\cite{Evlyukhin_NC_2014}. In picture (b) red nanoparticles are amorphized,
while yellow are annealed and crystalline.}
\label{fig:Fabr_2}
\end{figure}

In order to improve the location control of fabricated nanoparticles, laser ablation by focused beam can be used. Indeed, an
ultrashort laser pulse focused on silicon surface can heat material up to the critical point, leading to material fragmentation into
spherical nanoparticles and their deposition nearby the heated area~\cite{KuznetsovSciRep2012, Lukyanchuk13}
[Fig.~\ref{fig:Fabr_2}a]. It worth noting that colloids of chemically pure nanoparticles can be obtained by means of laser ablation
as well as the chemical deposition. The main advantages of the ablation approach are high-productivity and lack of harmful chemical
waste.

Fabrication of silicon nanoparticles, demonstrating Mie-resonances in visible range, with their localization control was carried out
by femtosecond laser focusing onto silicon surface, emitting nanoparticles to the transparent receiver
substrate~\cite{EvlyukhinNL2012,ZywietzAPA2014, Evlyukhin_NC_2014} [Fig.~\ref{fig:Fabr_2}b]. There are three main parameters
affecting ablated silicon nanoparticles: laser intensity, beam spatial distribution and sample thickness. For instance, single
silicon nanoparticle with a certain size can be formed from bulk silicon after irradiation by single laser pulse with ring-type
spatial distribution~\cite{ZywietzAPA2014}, or after irradiation of thin silicon film by conventional Gaussian
beam~\cite{Evlyukhin_NC_2014}.
Ultrashort laser can be used not only for fabrication, but also for silicon nanoparticles postprocessing. In particular, well-known
effect of laser annealing was applied for silicon nanoparticles to controllably change them from initially amorphized state to
crystalline one, tailoring their optical properties [Fig.~\ref{fig:Fabr_2}b]~\cite{Evlyukhin_NC_2014}.

\begin{figure}[!b] \centering
\includegraphics[width=0.5\columnwidth]{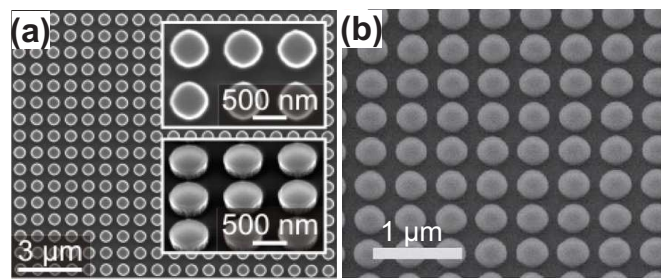}
\caption{Electron microscopy image of silicon nanoparticles obtained by means of reactive-ion-etching through a mask
(b)~\cite{StaudeACSNano2013} with additionally deposited Si$_3$N$_4$ thin film (b)~\cite{SpinelliNC2012}.}
\label{fig:Fabr_3}
\end{figure}

The most controllable fabrication of silicon nanoparticles was achieved by multi-stage method, including electron-beam lithography on
silicon-on-insulator wafers (formation of mask from resist) and reactive-ion-etching process with following removing of remaining
electron-beam resist mask. This advanced technology enables to form silicon nanocylinders [Fig.~\ref{fig:Fabr_3}b], in which
Mie-resonances can be precisely tuned by varying the basic geometrical parameters (diameter and height). Various types of structures
based on the silicon nanocylinders have been designed in order to show unique properties of the all-dielectric nanophotonics
devices~\cite{SpinelliNC2012, PersonNL2013, StaudeACSNano2013, ShcherbakovNL2014}. To achieve higher absorbtion of the fabricated
silicon metasurface, this method was supplemented by deposition of Si$_3$N$_4$ thin film~\cite{SpinelliNC2012}
[Fig.~\ref{fig:Fabr_3}b]. However, such lithography-based methods have such serious disadvantages as high-cost and low-productivity
of technological process in comparison with above mentioned lithography-free methods.

In order to summarize the section on methods of high-index dielectric (basically silicon) nanoparticles fabrication, we want to
stress here that currently developing approaches allow one to create various types of all-dielectric functional structures with given
optical properties, which will be discussed.

\section{Metadevices based on dielectric nanoparticles}

\subsection{Metamaterials and metasurfaces}
\label{sec:metamaterials}

Future technologies will push for a steep increase in photonic integration and energy efficiency, far surpassing that of bulk optical
components, silicon photonics, and plasmonic circuits. Such level of integration can be achieved by embedding the data processing and
waveguiding functionalities at the level of material rather than a chip, and the only possible solution to meet those challenges is
to employ the recently-emerged concept of \textit{metamaterials and metasurfaces}. Matamaterials are artificial media with exotic
electromagnetic properties not available in natural media which are especially created in order to reach functionalities required for
particular applications~\cite{Engheta2014,KivsharNM2012}. Metasurfaces are their two-dimensional implementations~\cite{Shalaev_2011}.
Metamaterials have been studied since 2000 and revealed such effects as negative refraction, backward waves, beating of
diffraction limit (subwavelength imaging), and became a paradigm for engineering electromagnetic space and controlling propagation of
waves by means of transformation optics~\cite{Engheta2014, Shalaev_2011, ZayatsNP2012}. The research agenda is now focusing on
achieving tuneable, switchable, nonlinear and sensing functionalities of metamaterials. Since 2010 the studies have been shifted to
the stage of practical implementation and development of real matadevices. As a result, a novel concept of metadevices, that can be defined as metamaterial-based devices with novel and useful functionalities achieved by structuring of functional matter on the subwavelength scale, has being developed.~\cite{KivsharNM2012}. The metadevices practical implementation is the general
trend in the area of metamaterials.

The area of nanophotonics metamaterials have opened a broad range of technologically important capabilities ranging from the subwavelength focusing
to “stopped light”, including their ability to control magnetic response of novel subwavelength structured materials. This is important
because the magnetic response of natural materials at optical frequencies is very weak due to diminishing of  electronic spin states at high frequencies~\cite{Landau_CM}. That is why only the electric component of light is directly controlled in
conventional photonic devices. However, effective control of light at the nanoscale requires presence of electric and magnetic
response, simultaneously. A vast majority of the current metamaterial structures exhibiting magnetic response contain metallic
elements with high conductive losses at optical frequencies, which limits their performance. One of the canonical examples is a
split-ring resonator (SRR) that is an inductive metallic ring with a gap that is a building block of many metamaterials. The SRR
concept works very well for gigahertz~\cite{Smith, Shelby}, terahertz~\cite{Padilla06} and even near-infrared (few hundreds
THz)~\cite{Liu08} frequencies. However, for shorter wavelengths and in particular for visible spectral range this concept fails due
to increasing losses and technological difficulties to fabricate smaller and smaller constituting split-ring
elements~\cite{Soukoulis07}.
\begin{figure}[!t] \centering
\includegraphics[width=0.7\columnwidth]{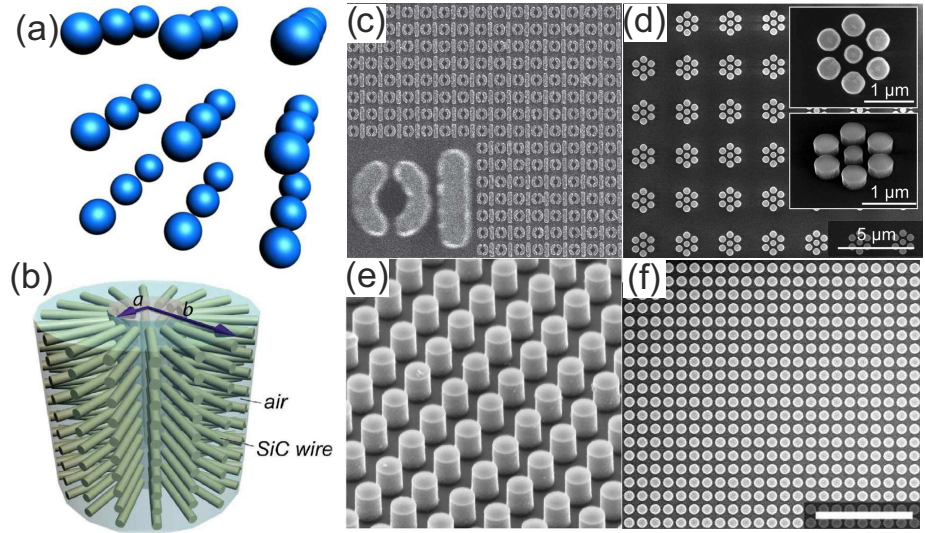}
\caption{(a,b) All-dielectric metamaterials based on spherical and cylindrical particles~\cite{Mosallaei2008, Zhao09, Odit2009,
TretyakovJAP2006}; (c-f) all-dielectric metasurfaces~\cite{Staude_15, SpinelliNC2012, Valentine2014}.}
\label{fig:meta}
\end{figure}

In order to overcome these fumdamental problems, an altermative approach of all-dielectric metamaterials has been
proposed~\cite{Grzegorczyk2007, ShullerPRL2007, Meng2008, Mosallaei2008, Guizal2009, Zhao09, Cummer_08, Brener_12, Lukyanchuk13,
Zhou2014, Valentine2014, Limonov2015}. In this case, a high-index dielectric particle, exhibiting magnetic and electric Mie resonances, plays role of a single meta-atom. Such high-index dielectric particles can replace their metallic counterparts in metamaterials
and metasurfaces due to the absence of free charges. For example, it has been shown that 3D dielectric composite of high-index
dielectric particles (as shown in Fig.~\ref{fig:meta}a) exhibits negative permeability near the first Mie resonance~\cite{Meng2008,
Mosallaei2008}. For a composite consisting of high-index dielectric spherical particles embedded in a low-index dielectric matrix, the relative
effective permeability can be given according to Levin’s mode by~\cite{Lewin_1947, Mosallaei2008}
\begin{equation}\label{Lewin}
    \mu_{\rm ef}=1+\frac{3u}{(F(\theta)+2)/(F(\theta)-1)-u},
\end{equation}
where
\begin{equation}\label{LF}
F(\theta)=\frac{2(\sin\theta-\theta\cos\theta)}{(\theta^2-1)\sin\theta+\theta\cos\theta},
\end{equation}
the volume fraction of the spherical particles $u=4/3\pi(R_s/a)^3$, $\theta=k_0 R_s \sqrt{\varepsilon}$, and where $R_s$ and $a$ are
the particle radius and lattice constant, $k_0$ and $\varepsilon$ are the free-space wavenumber and permittivity of particles,
respectively. Even more complex geometry of such all-dielectric metamaterials have been proposed in works~\cite{Odit2009,
TretyakovJAP2006}.

Conventional optical components rely on gradual phase shifts accumulated during light
propagation to shape light beams. The nanostructured design can introduce new degrees of freedom by making abrupt phase changes over
the scale of the wavelength. A two-dimensional lattice of optical resonators or nanoantennas on a planar surface, with spatially
varying phase response and subwavelength separation, can imprint such phase patterns and discontinuities on propagating light as it
traverses the interface between two media. In this regime, anomalous reflection and refraction phenomena can be observed in optically
thin metamaterial layers, or optical metasurfaces, creating surfaces with unique functionalities and engineered reflection and
transmission laws. The first example of such a metasurface, being a lattice of metallic nanoantennas on silicon with a linear
phase variation along the interface, was demonstrated recently~\cite{Gaburro11, Zhang12NL, Tsai12NL, AluPRL2013, Grbic2013}. The
concept of metasurfaces with phase discontinuities allows introducing generalized laws for light reflection, and such surfaces
provide great flexibility in the control of light beams, being also associated with the generation of optical vortices. Metasurfaces
can also be used for the implementation of important applications such as light bending ~\cite{Gaburro11} and specific
lenses~\cite{Capasso12}.

The gradient metasurfaces created by high-dielectric nanoparticles of varying shape have been recently proposed~\cite{Staude_15,
Capasso_2015}. In the article~\cite{Staude_15}, for the first time, highly efficient all-dielectric metasurfaces for near-infrared
frequencies using arrays of silicon nanodisks as meta-atoms have been demonstrated (see Fig.~\ref{fig:meta}f). The authors have
employed the main features of Huygens' sources, namely spectrally overlapping electric and magnetic dipole resonances of equal
strength, to demonstrate Huygens' metasurfaces with a full transmission-phase coverage of 360 degrees and near-unity transmission,
and confirmed experimentally full phase coverage combined with high efficiency in transmission. Based on these key properties, the
authors show that all-dielectric Huygens' metasurfaces could become a new paradigm for flat optical devices, including beam-steering,
beam-shaping, and focusing, as well as holography and dispersion control.

\subsection{Coupled-resonator optical waveguides}
\label{sec:waveguides}

A design of highly efficient integrated circuits with combined optical and electronic components for the subwavelength guiding of the
electromagnetic energy is one of the main trends of the modern nanophotonics~\cite{ChenSciRep2012}. In order to achieve high
integration densities, optical waveguides with subwavelength light localization have been proposed. Conventional silicon wire
waveguides have a minimal cross-section size, and they can be manufactured being of a high quality~\cite{LawScience2004}. However,
such waveguides do not provide low-loss propagation of optical signals through sharp bends, and require rather large bending
geometries thus increasing the overall size of an optical chip~\cite{EspinolaOE2001}. Photonic crystals have been viewed as a
possible alternative, and it has been already demonstrated that light can be guided by a waveguide composed of defects, and such
waveguides can have sharp bends~\cite{MiroshnichenkoOE2005}. However, this nice property of photonic crystals to give light though
sharp bends was found to depend strongly on the bend geometry being also linked to the strict resonant conditions associated with the
Fano resonance where the waveguide bend plays a role of a specific localized defect~\cite{MiroshnichenkoOE2005}, thus demonstrating
narrowband functionalities.

Another candidate for the efficient propagation through sharp bends is a coupled-resonator optical waveguide
(CROW)~\cite{YarivOL1999}. Such waveguides can be realized as chains of metallic nanoparticles guiding the electromagnetic energy via
plasmonic resonances~\cite{AluPRB2006}. Theoretical studies in the quasistatic single-dipole approximation suggested an efficient
power transmission through sharp bends in such waveguides by converting the mode polarization~\cite{MaierAdvMat2001}. Small size of
nanoparticles (several tens of nanometers) made these waveguides very attractive for nanophotonic applications. However, later
studies revealed that longitudinally and transversely polarized modes split in the frequency as soon as nonquasistatic approximation
is considered~\cite{WeberPRB2004}, providing several practical limitations for the polarization conversion at the bends in chains of
plasmonic nanoparticles. Moreover, due to strong Ohmic losses, the light propagation in plasmonic chains is strongly limited by short
distances~\cite{SolisNanoLetters2012}.

The CROW-type structures with the subwavelength guiding and low losses have been demonstrated with arrays of dielectric nanoparticles
with high values of the refractive index~\cite{SavelevPRB2014}. Dielectric nanoparticles support both magnetic dipole (MD) and
electric dipole (ED) resonances simultaneously~\cite{KuznetsovSciRep2012, EvlyukhinNL2012}, which gives an additional control
parameter over the light scattering~\cite{KrasnokOE2012, BonodOE2012}, and the waveguides composed of such nanoparticles were shown
to support several modes of different types~\cite{SavelevPRB2014}. In the case of spherical particles, the transversely polarized
modes (with both MD and ED oriented perpendicular to the chain direction) and longitudinally polarized modes (when the dipoles are
oriented along the chain direction) correspond to the same resonance frequencies, and they form separate pass bands in different
spectral ranges, due to the different dipole-dipole interaction. For non-spherical particles, resonance frequencies depend on the
orientation of dipoles, and the corresponding pass bands can be shifted by changing the particle parameters.

\begin{figure*}[!t]\centering
\includegraphics[width=0.7\columnwidth]{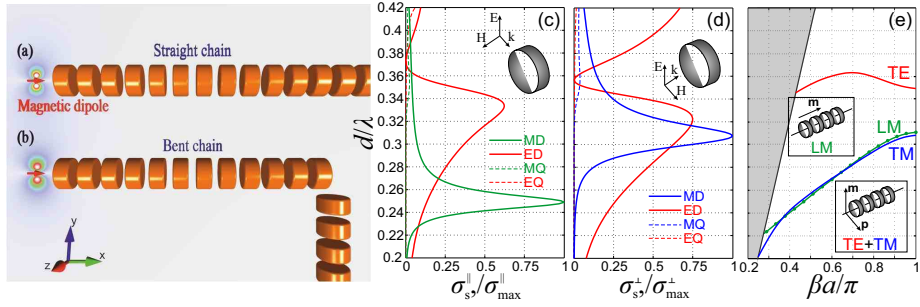}
\caption{Geometry of two types of all-dielectric waveguiding structures: (a) straight waveguide, and
(b) waveguide with a sharp bend. A magnetic dipole is employed as a source to excite the guided modes. (c,d) Multipole decompositions
of the scattering spectra of a dielectric disk irradiated by a plane wave; the orientations of external wave polarization are shown
in the insets. (e) Dispersion diagram for an infinite chain of side located lossless nanodisks. Here $d/\lambda$ is the normalized
frequency; $d$ is the diameter of a nanodisk, and $\lambda$ is the wavelength.}
\label{fig:cyl+chain}
\end{figure*}

In the Ref.~\cite{Savelev2014_1} the transmission efficiency of the CROW-type waveguides composed of arrays of high-index dielectric
nanodisks with and without sharp bends have been studied. Such structure is shown in Figs.~\ref{fig:cyl+chain}a,b. Such waveguides
can be highly tunable, because they have several independent parameters: $h$ and $d$ (height and diameter of the nanodisk,
respectively) and $a$ (period of the chain). The appropriate period of the chain has been chosen so the longitudinal and transverse
magnetic pass bands in the nanodisk chain overlap (which cannot be done with spherical particles). This condition provides an
efficient transmission through sharp 90$^\circ$ bends. In this article the theoretical conclusions have been supported by presenting
the experimental results for the microwave frequencies for an efficient guiding through 90$^\circ$ bend in a microwave dielectric
waveguide. The multipole moments of a nanodisk depend on the incident wave direction, and the simulated multipole expansion of the
nanodisk scattering spectra for both side and top incidence are presented in Figs.~\ref{fig:cyl+chain}(c,d), respectively. In the
case of the side incidence, the wave vector is oriented along the nanodisk axis, whereas for the top incidence the wave vector is
perpendicular to the nanodisk axis (see the insets). For the nanodisk under consideration, the first resonance is MD resonance, and
the second resonance is the ED resonance. For the side incidence [see Fig.~\ref{fig:cyl+chain}c], the MD and ED resonances are
separated. The MD is observed at the lower frequencies $d/\lambda \approx 0.25$, due to larger size of the effective cross section of
the particle. For the top incidence, the MD and ED resonances approximately overlap when $d/\lambda \approx 0.31$ [see
Fig.~\ref{fig:cyl+chain}d], which is in full agreement with previous numerical calculations~\cite{EvlyukhinPRB2011}. The higher order
multipole moments are significant only for higher frequencies $d/\lambda \gtrsim 0.45$. Figure~\ref{fig:cyl+chain}(e) summarizes the
waveguide modes of an infinite straight chain of nanodisks simulated by the eigenmode solver in CST Microwave Studio. Periodic
boundary conditions have been applied in the $x$ direction, the electric boundary conditions have been used in both $y$ and $z$
directions. Here (i) longitudinal magnetic (LM) mode that corresponds to the MD resonance in Fig.~\ref{fig:cyl+chain}a, (ii)
transverse magnetic (TM) mode that corresponds to MD resonance in Fig.~\ref{fig:cyl+chain}c, and (iii) transverse electric (TE) mode
that corresponds to the ED resonance in Fig.~\ref{fig:cyl+chain}d are presented. Other modes fall out of the considered spectral
region, and they are not shown in Fig.~\ref{fig:cyl+chain}c. By choosing the appropriate period of the chain ($a=1.25 h$) the LM and
TM modes overlap can be achieved. This provides an opportunity for efficient light transmittance in a chain with sharp bends.

\begin{figure}[!b]\centering
\includegraphics[width=0.5\columnwidth]{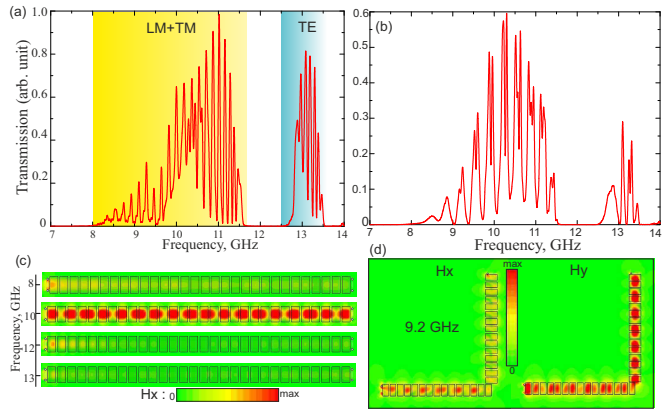}
\caption{The results of experimental measurements of normalized power transmittance of a straight chain waveguide composed of 30
ceramic cylinders (a) and a chain with a 90 $^\circ$ bend (b). (c,d) Distribution of the $x$- and $y$-components of the magnetic
field for different frequencies.}
\label{fig:straight_chain}
\end{figure}

The transmission efficiency is shown in Fig.~\ref{fig:straight_chain}a. The transmission efficiency is defined as
$|S_{21}|^2/|S_{21}|^2_\mathrm{max}$, were $|S_{21}|^2$ is the power transmitted from the first to second port normalized to the
maximum value $|S_{21}|^2_\mathrm{max}$. Accordingly to the numerical results for a finite chain, the first pass-band is observed at
8--11.6~GHz (normalized frequencies 0.21 -- 0.3 $d/\lambda$), which is in the total agreement with this theoretical prediction and
the dispersion dependence [see Fig.~\ref{fig:cyl+chain}e]. The pass-band can be explained by the presence of Fabry-Perot resonances
of both LM and TM modes in a finite chain. The second pass-band corresponds to the TE mode of Fig.~\ref{fig:cyl+chain}e, and it
appears in the frequency range 12.6--13.6~GHz (normalized frequencies 0.33 -- 0.36 $d/\lambda$). The magnetic field distribution in
the straight waveguide at two characteristic frequencies is shown in Figs.~\ref{fig:straight_chain}c. Magnitude of the longitudinal
($H_x$) and transverse ($H_y$) components of the magnetic field at the lower edge of the first pass-band (8~GHz) are negligibly
small, and therefore no transmission is observed. In the pass-band where both LM and TM modes exist simultaneously, the
high-efficient transmission is observed. The frequency range 11.6--12.6~GHz corresponds to the stop-band, and therefore no
transmission is observed. In the second pass-band where only TE modes propagate, the magnitude of the longitudinal component of the
magnetic field is close to zero, and for the demonstration the fields at 12~GHz is shown.

Then an all-dielectric waveguide composed of 30 ceramic disks with a bend in the middle has been studied (see
Fig.~\ref{fig:straight_chain}b,d). The maximum value of the transmission efficiency of the waveguide with a band is of the level of
0.6. The losses caused by the bend depend on the frequency, and they do not exceed 2.2 dB. The efficient transmission through the 90
$^\circ$ bend is achieved due to an overlap of the LM and TM modes. As follows from Figs.~\ref{fig:straight_chain}b,d the LM mode in
the horizontal branch transforms into the TM mode propagating in the vertical branch and vice versa.

Thus, the discrete waveguides based on high-index dielectric nanoparticles may exceed its currently existing analogs: plasmon
waveguides, dielectric photonic crystals and homogeneous silicon waveguides, due to the large number of customizable options and
negligible energy dissipation. Therefore, these waveguides can be used in photonic components responsible for the transmission of
information in the optical and optoelectronic integrated circuits.

\subsection{Nanoantennas and oligomers}
\label{sec:nanoantennas}

The recently emerged field of optical nanoantennas is promising for its potential applications in various areas of nanotechnology. The ability to redirect propagating radiation and transfer it into localized subwavelength modes at the nanoscale~\cite{Novotny_10_NatPhot} makes the optical nanoantennas highly desirable for many applications. Originally, antennas were suggested as sources of electromagnetic radiation at radio frequencies and microwaves, emitting radiation via oscillating currents. Different types of antennas were suggested and demonstrated for the effective manipulation of the electromagnetic radiation~\cite{Balanis}. Thus, conventional antennas perform a twofold function as a source and transformation of electromagnetic radiation, resulting in their sizes being comparable with the operational wavelength. Recent success in the fabrication of nanoscale elements allows to bring the concept of the radio frequency antennas to optics, leading to the development of optical nanoantennas consisting of subwavelength elements~\cite{Novotny_10_NatPhot}. Currently nanoantennas are used for near-field microscopy~\cite{Fan:NL:2012}, high resolution biomedical sensors~\cite{Zhang:NL:2011}, photovoltaics~\cite{Spinelli2012}, and medicine~\cite{rac_10}. This section is devoted to a review of nanoantennas based only on dielectric nanoparticles witch have been suggested in Refs.~\cite{Krasnok2011,
Krasnok_APL_12, KrasnokOE2012}.

\begin{figure}[!b]\centering
\includegraphics[width=0.5\textwidth]{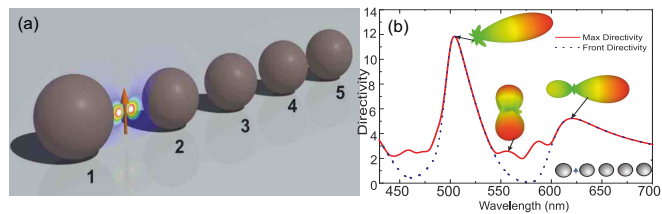}
\caption{(a) Dielectric optical Yagi-Uda nanoantenna, consisting of the reflector of the radius $R_r=75$ nm, and smaller director of
the radii $R_d=70$ nm. The dipole source is placed equally from the reflector and the first director surfaces at the distance G. The
separation between surfaces of the neighbouring directors is also equal to G. (b) Directivity of the dielectric Yagi-Uda nanoantenna
vs wavelength for the separation distance $G=70$ nm. Insert demonstrates 3D radiation pattern diagrams at particular wavelengths.}
\label{fig1}
\end{figure}

This novel type of optical nanoantennas made of all-dielectric elements and can be considered as the best alternative to their
metallic counterparts. First, dielectric materials exhibit low loss at the optical frequencies. Second, as was suggested
earlier~\cite{KuznetsovSciRep2012}, nanoparticles made of high-index dielectrics (silicon) may support both electric and magnetic
resonant modes. This feature may greatly expand the applicability of optical nanoantennas for, e.g. detection of magnetic dipole
transitions of molecules. The real part of the permittivity of the silicon is about $18$~\cite{VuyeSi}, while the imaginary part is
up to two orders of magnitude smaller than that of nobel metals (silver and gold). Then, a novel concept of superdirective
nanoantennas based on the generation of higher-order optically-induced magnetic multipoles have been
introduced~\cite{KrasnokNanoscale, KrasnokAPLSuperdir}. Such an all-dielectric nanoantenna can be realized as an optically small
spherical dielectric nanoparticle with a notch excited by a point source located in the notch. Proposed superdirectivity effect is
not associated with high dissipative losses, because of the magnetic nature of the nanoantenna operation.

The mentioned above properties of dielectric nanoparticles allow to realize optical analogue of the
Yagi-Uda design (see Fig.\ref{fig1}a)~\cite{Balanis} consisting of four directors (dielectric nanoparticles) and one reflector and
point-like electric dipole. The radii of the directors and the reflector are chosen to achieve the maximal constructive interference
in the forward direction along the array. The optimal performance of the Yagi-Uda nanoantenna should be expected when the radii of
the directors correspond to the magnetic resonance, and the radius of the reflector correspond to the electric resonance at a given
frequency, with the coupling between the elements taken into account. This particular design consists of the directors with radii
$R_d=70$ nm and the reflector with the radius $R_r=75$ nm. In Fig.~\ref{fig1}b the directivity of all-dielectric Yagi-Uda nanoantenna
vs. wavelength with the separation distance $D=70$ nm is presented. Inserts demonstrate the 3D radiation patterns at particular
wavelengths. A strong maximum at $\lambda = 500$ nm have been achieved. The main lobe is extremely narrow with the beam-width about
$40^\circ$ and negligible backscattering. The maximum does not correspond exactly to either magnetic or electric resonances of a
single dielectric sphere, which implies the importance of the interaction between constitutive nanoparticles.

\begin{figure}[!t] \centering
\includegraphics[width=0.5\textwidth]{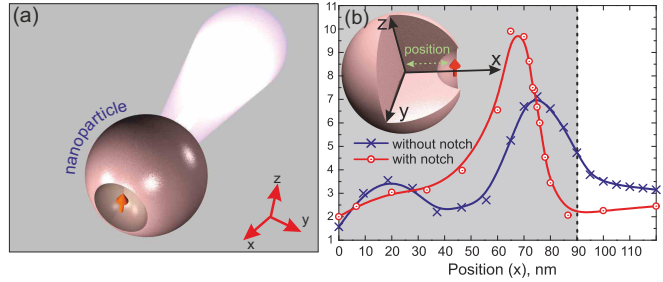}
\caption{(a) Geometry of the notched all-dielectric nanoantenna. (b) Maximum of directivity depending on the position of the dipole
($\lambda=455$ nm) in the case of a sphere with and without notch, respectively.} \label{fig2}
\end{figure}

Electrically small radiating systems whose directivity exceeds significantly that
of a dipole are usually called superdirective~\cite{Balanis}. Achieving high radiation directivity is very important for actively
studied optical nanoantennas~\cite{AluWireless, Novotny_10_NatPhot}. However antenna's superdirectivity in the optical frequency
range was not discussed or demonstrated so far. In the article~\cite{KrasnokNanoscale} the novel concept of superdirective
nanoantennas based on the excitation of higher-order magnetic multipole moments in subwavelength dielectric nanoparticle have been
proposed. The superdirective regime have been achieved by placing a nanoemitter (e.g. a quantum dot) within a small notch created on
the dielectric nanosphere surface, as shown in Fig.~\ref{fig2}a. The notch has the shape of a hemisphere with a radius $R_{\rm n}\ll
R_{\rm s}$. The emitter can be modeled as a point-like dipole and it is shown in the figure by a red arrow. It turns out that such a
small modification of the sphere would allow the efficient excitation of higher-order spherical multipole modes. Figure~\ref{fig2}b
shows the dependence of the maximum directivity $D_{\rm mav}$ on the position of the emitting dipole in the case of a sphere $R_{\rm
s}=90$~nm without a notch, at the wavelength $\lambda=455$~nm (blue curve with crosses). This dependence has the maximum ($D_{\rm
max}=7.1$) when the emitter is placed inside the particle at the distance 20~nm from its surface. The analysis shows that in this
case the electric field distribution inside a particle corresponds to the noticeable excitation of higher-order multipole modes. This
becomes possible due to strong inhomogeneity of the external field produced by the nanoemitter. Furthermore, the excitation of
higher-order multipoles can be significantly improved by making a small notch in the silicon spherical nanoparticle and placing the
emitter inside that notch, as shown in Fig.~\ref{fig2}a. This modification of the nanoparticle transforms it into a resonator for
high-order multipole moments. The notch has the form of a hemisphere with the center it the dielectric nanoparticle's surface. The
optimal radius of the notch is $R_{\rm n}=40$ nm, that have been found by means of numerical optimization. Red curve with circles in
the Fig.~\ref{fig2}b shows maximum of directivity corresponding to this geometry. Maximal directivity at wavelength 455~nm is $D_{\rm
max}=10$. Likewise, the absence of hot spots inside the nanoantenna leads to low dissipation in the radiation regime, so that this
dielectric nanoantenna has significantly smaller losses and high radiation efficiency of up to 70$\%$.

Fano resonance~\cite{Fano_61,Miroshnichenko:NL:2012,Liu2015} is known to originate from interference of two scattering
channels, one of which is non-resonant, and another is strongly resonant. Fano resonance was observed in different branches of
physics, including photonics, plasmonics and metamaterials~\cite{Lukyanchuk:NM:2010}. It is highly sensitive to the optical
properties of the background medium, which makes it very perspective in design of sensors.

In the last few years there is a growing interest in studying Fano resonances in the the so-called plasmonic \textit{oligomer
structures}, that consist of several symmetrically positioned metallic nanoparticles~\cite{Rahmani:NL:2012, Zhang:PNAS:2013}. In such
structures Fano resonance appears as a resonance suppression of the cross-section of the structure, and it is accompanied by strong
absorption. This suppression can be explained as a result of destructive interference of the two excited plasmonic modes in the
strucure, one of which is resonant.

\begin{figure}[!t]\centering
\includegraphics[width=0.45\columnwidth]{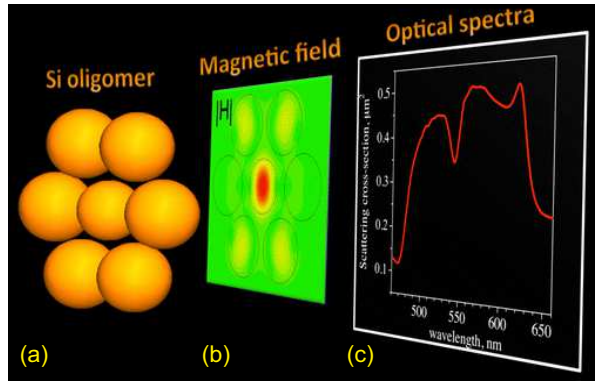}
\caption{(a) Geometry of dieletric oligomer that consists of six identical nanoparticles, positioned in the vertices of the regular
hexagon, and the particle of smaller radius in the center. (b) Distribution of the magnetic field intensity in the oligomer at the
Fano resonance wavelength. (c) Spectral dependence of the scattering cross-section of the oligomer; the dip at at the Fano resonance
wavelength can be observed (about 550 nm). From Refs.~\cite{Miroshnichenko:NL:2012,Filonov_oligomer}.}\label{oligomer}
\end{figure}

Recently it was shown that dielectric oligomers are also able to exhibit Fano resonance~\cite{Miroshnichenko:NL:2012,
Staude_2014Small, Filonov_oligomer}. The important feature of dielectric oligomers, comparing to their metallic counterparts, is the
localization of electromagnetic field inside dielectric nanoparticles.
Another important property of such oligomers is that the fudamental mode of a basic element of such oligomers -- high-index spherical
nanoparticle -- is a magnetic dipole mode~\cite{KuznetsovSciRep2012, EvlyukhinNL2012}. Formation of this magnetic mode, as was
mentioned earlier, is due to excitation of circulating displacement current. It occurs, when the diameter of the particle is
comparable to the wavelength inside the nanoparticle.

Authors of the Ref.~\cite{Miroshnichenko:NL:2012} have shown that the structure that consists of six identical dielectric
nanoparticles, positioned in the vertices of the regular hexagon, and the particle of another radius in the center [see
Fig.~\ref{oligomer}a], exhibits Fano resonance at the resonance frequency of the central particle, while six other particles do not
resonate at this frequency and they form a non-resonant mode of the whole structure. Near-field interference of this two modes leads
to the suppression of the whole structure scattering~\cite{Miroshnichenko:NL:2012}.

In the paper~\cite{Filonov_oligomer} authors experimentally proved the existence of the Fano resonances in dielectric oligomers. Due to the scalability of Maxwell equations authors used macroscopic
ceramic spheres with sizes of several centimeters (instead of silicon nanoparticles). Such particles exhibit magnetic response in microwave frequency range. Authors measured magnetic field in the vicinity of
dielectric oligomer with high accuracy, which allowed to verify the origin of Fano resonance, predicted in the theoretical study~\cite{Miroshnichenko:NL:2012} [see Fig.~\ref{oligomer}b]. The full cross-section of oligomer structures did not depend on polarization of the incident wave~\cite{Hopkins:N:2013}. Theoretically such structures were described with discrete dipole approach in good agreement with experimental results.

\section{Conclusions and outlook}

We have reviewed very briefly some of the recent developments in the field of dielectric nanophotonics. This branch of optical science studies light interaction with high-index dielectric nanoparticles supporting optically-induced electric and magnetic Mie resonances.
We have described several advances in this field which demonstrate that dielectric structures allow to control both magnetic and electric components of light in a desirable way, and also discuss properties of high-indexed nanoparticles along with their fabrication methods. We have discussed typical examples of dielectric structures such as discrete waveguides, dielectric nanoantennas, oligomers and metamaterials.

Importantly, future technologies will demand a huge increase in photonic integration and energy efficiency far surpassing that of bulk optical components and silicon photonics. Such a integration can be achieved by embedding the data-processing and waveguiding functionalities at the material’s level, creating the new paradigm of metadevices. It is now believed that robust and reliable metadevices will allow photonics to compete with electronics not only in telecommunication systems, but also at the level of consumer products such as mobile phones or automobiles. The main challenges in achieving this vision will be in developing cost-efficient fabrication and device integration technologies. All-dielectric nanophotonics is seen as a practical way to implement many of the important concepts of metamaterials allowing high functionalities and low-loss performance of metadevices.


\end{document}